\documentstyle[epsf]{aipproc}
\newcommand{\be}{\begin{eqnarray}}

\newcommand{\ee}{\end{eqnarray}}

\title{
        \begin{flushright}
        {\normalsize
        TPI-MINN-99-13\\
        NUC-MINN-99/5-T\\
        UMN-TH-1749\\
        March 1999 \\}
        \end{flushright}
\bf     Initial Conditions for Parton Cascades\footnote{Talk presented at
RHIC Physics and Beyond: Kay Kay Gee Day, Brookhaven National Laboratory, Upton,
Long Island NY, Oct. 1998}      
       }
\author{Larry McLerran\\
       {\small\it Theoretical Physics Institute, University of
Minnesota,
        Minneapolis, MN 55455  } \\         
       }

\date{}

\begin{document}
\maketitle

\begin{abstract}
Abstract:  
I discuss the initial conditions for a parton cascade.
\end{abstract}

\section{Introduction}

Klaus Kinder-Geiger was a postdoctoral fellow with us at the University of
Minnesota from 1991-1993.  

I remember well the first seminar he gave to us on
the work he had been doing with Berndt Muller concerning the parton cascade
model. I was very excited by what he was doing, and I was asking him question
after question.  This was at a time at the University of Minnesota recently
after we had hired a number of Russians, so 2 hour seminars were not unusual.  
(We no longer have such long seminars.  The Russians are real Americans now.)
Klaus was needless to say nervous about the time he was taking, and Sharon,
who was in the audience, was I think more than a little bothered by what she
perceived as my harassment of Klaus.  

I don't think Klaus fully understood how much I respected him after that talk.
He was one of the few young people I knew who were not afraid to say they
didn't understand something, who was excited about exploring new ways of
thinking, and most important, had a deep understanding of what it was he was
doing.

I guess we all knew Klaus as an unconventional thinker.  When he was with us,
he had the typical fears and lack of confidence of all people his age.  The
following story illustrates this:  We were discussing hiring new postdocs at
Minnesota and Joe Kapusta and I invited Klaus to come to the meeting and join
in the discussion.  The first thing I did was go down the list of people, and
anyone who had published less than three papers a year since they got their
Ph. D., I refused to further consider.  Klaus muttered something about how
this wasn't really very fair, and I answered back that I didn't want to hire
lazy people.  Klaus had been with us for about 6 months at this time and had
submitted a paper, maybe two, for publication.  In the next few months, he
submitted about half a dozen.

Klaus was the most prolific postdoc which we ever hired in nuclear theory at
the University of Minnesota.  His papers were not superficial and each
involved much work and thinking.  

Klaus and I talked much but never worked
together.  He was captured by my colleague Joe Kapusta.  Klaus had a profound
impact on my thinking nevertheless.  He got me very interested in his picture
of the very early stages of heavy ion collisions, and this is the subject of
this talk.

To understand the parton cascade model,\cite{gm}-\cite{g} one needs a
space-time picture of
nucleus-nucleus collisions.  Such a space-time picture was developed by Bjorken
and I shall summarize it in this introduction.\cite{bjorken} We concentrate on
the central region of collisions at asymptotically high energy.  We assume
that the rapidity density of produced particle is slowly varying, slow enough
so that we can treat the distribution
\be
               {{dN} \over {dy}} = contant
\ee
If this is the case, the space-time dynamics for particle produced in the
central region should be longitudinally Lorentz boost invariant.  This means
that the dynamical evolution of the particles produced in the collision is
described by only on parameter $\tau = \sqrt{t^2-z^2}$.  We also will assume
that the transverse size of the system is large enough so that one can ignore
effects such as the transverse expansion of the system.  The other
longitudinal variable is the space-time rapidity
\be
        \eta = {1 \over 2} ln\left( {{t+z} \over {t-z}} \right)
\ee
which under a longitudinal Lorentz boost changes by a constant.

Note that for a free streaming particle, $\eta = y$ since
\be
        \eta = {1\over 2} ln \left( {{t+z} \over {t-z}} \right) =
{1\over 2} ln \left( {{1+v_z} \over {1-v_z}} \right) =
{1 \over 2} ln \left( {{E+p_z} \over {E-p_z}} \right) = y
\ee
We see therefore that
\be
        {{dN} \over {dy}} \sim {{dN} \over {d\eta}} = {{dN} \over {dz}}
{t \over\tau^2}
\ee
The particles produced at $z = y = 0$ therefore expand and dilute their
density
as 
\be
        {{dN} \over {dz}} = {constant \over t}
\ee
In an isentropic expansion, as will be the case later in the collision after
the particle have thermalize, the entropy density $\sigma $ satisfies
\be
        \tau \sigma = \tau_0 \sigma_0
\ee
Here $\tau_0$ is the initial thermalization time, for which a variety of
arguments suggest that $\tau_0 \le 1 Fm/c$

We expect that the entropy will be approximately conserved as the system
expands.  If there is a first order phase transition, then there will be some
entropy production, but again for the typical time scales characteristic of
heavy ion collisions, we do not expect a dramatic increase in the entropy.
Of course as the system expands, the degrees of freedom of the system change
dramatically.  Early on, we expect that the system will be an almost ideal gas
of quarks and gluons.  Late on the gas is hadronic, and very late it is a gas
of far separated almost non-interacting pions. If the hadronic gas decouples
when the pions are still to a good approximation massless, as will be the case
for decoupling temperatures $T_{decoupling} \ge 100 MeV$ (recall the average
energy is 3T, so that $(m_{pion}/E)^2 \sim .1$), then one can show
that entropy production implies that the number of gluons initially is the same
as the number of pions.  Therefore the number density of gluons early on is
\be
        {N \over V} = {1 \over {\tau_0 \pi R^2}} {{dN_{pions}} \over {dy}}
\ee
For the typical rapidity density of pions seen at RHIC, this leads to initial
temperatures $T \ge 200 MeV$.  This should be sufficient to produce a quark
gluon plasma.   

In Fig. 1, a space-time picture of the evolution of matter produced in
ultra-relativistic nuclear collisions is shown.  After the time $\tau_0$ when
thermalization occurs, the system expands.  At some time and corresponding
temperature, the system converts from a quark-gluon plasma into a hadron gas.
This may take some time, and go through a mixed phase if there is truly a
phase difference between hadronic matter and a quark-gluon plasma.  If there
is no true phase change, the system nevertheless changes its properties 
dramatically and to do so involves time.  At much later time, the system
freezes out and produces free streaming particles.

\begin{figure}
\begin{center}
\epsfxsize=10cm
\leavevmode
\hbox{ \epsffile{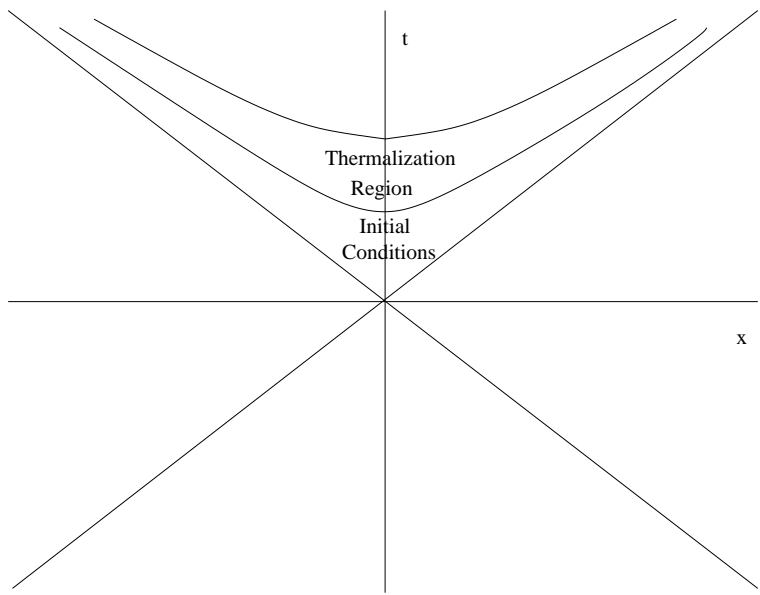}}
\end{center}
\caption{A Space-Time Diagram of Ultrarelativistic Nuclear Collisions}
\label{fig1}
\end{figure}

\section{What Happens Before $\tau_0$ ?}

What happens in the time between $\tau = 0$ and $\tau_0$?  
Surely, the earlier one goes in time, the more energetic are particle
interactions.  Weak coupling methods should therefore be at their best,
and one should be able to compute, at least in the limit of very high energies
and very large nuclei, from first principles in QCD.

It is reasonable to assume that the matter when it is first formed in heavy
ion collisions is in some sort of non-thermal distribution.  The matter must
therefore thermalize. Klaus Kinder-Geiger and Berndt Muller made a daring
proposal in an attempt to understand the thermalization.  They assumed the
momentum space-distribution of partons just after formation was given by the
parton distribution functions.  

This assumption deserves a little comment since the parton distribution
functions specify only the longitudinal momentum space distribution of the
partons.  Both the transverse momentum structure and the space-time positions
of the partons at formation must be assumed.  Some guidance about what are
reasonable assumptions are given by uncertainty principle arguments, but the
coordinate space picture is nevertheless assumed.

In fact the uncertainty principle and quantum mechanics limits the region
where the parton cascade can be applied.  In order to use a cascade, one must
specify the phase space distribution of particles $f(\vec{p},\vec{x},t)$.
This involves specifying both the position and coordinates of the particles,
and is inconsistent with a quantum mechanical description.  (One can formally
define a phase space distribution function for a fully quantum system, but the
distribution will in general lack positivity, and usually will violate it in
the region of phase space where the quantum effects are important.)  At the
earliest times in the collision, the system is described by two quantum
mechanical wavefunctions which describe the nuclei.  Therefore for some
sufficiently early time, the parton cascade description must fail.

One can also see that one must go beyond partons to describe the earliest
times in the collisions.  At the earliest time, the density of fast moving
quarks and gluons is very high.  If we use cascade theory to describe their
effect on long wavelength quanta such as will be produced in the central
region, we will have each of the quanta acting incoherently.  This is because
in a cascade, only matrix elements squared for single particle scattering
occur.  On the other hand, we know that when we compute the field associated
with these quanta, their effect is tempered since because of overall color
neutrality for confined particles. Any colored field will therefore be reduced
in strength in the infrared, and their effect on long wavelength quanta will
be reduced.

Klaus and Berndt tried to phenomenologically include the effect of quantum
mechanics and classical charge coherence in two separate ways.  The first was
to assume that particles were not produced and could interact until after a
characteristic formation time took place in the rest frame of the particle.
This has the effect of delaying the cascade description until after the
formation time has taken place.  It evades the question of whether the parton
distributions are modified during the time from the initial collision $\tau =
0$ until the formation time.  During this time, the evolution is quantum
mechanical, but in the complicated collision environment, there may be a
non-trivial quantum evolution of the distributions typical of a single nucleus.

The other way they tried to build in some coherence is in cutting of the cross
sections for parton parton scattering at small angles. In a plasma, for
example, such cross sections are cutoff by media screening effects.  This
parameter is crucial in their computations as all cross sections depend
quadratically on such a cutoff.

In spite of these difficulties, the parton cascade model provides a useful way
to describe the evolution of the matter from some time which I will refer to
as the formation time $\tau_f$ until the thermalization time.  The details of
what is the precise form of the initial conditions and how one cuts off cross
sections may be subject to dispute, but the description of the time evolution
between $\tau_f$ and $\tau_0$ is conceptually correct.

There are several qualitative issues associated with the approach to
equilibrium which we can easily understand from this approach.  The first is
that if the partons are formed in an energetic environment, then the coupling
is weak.  Thermalization will take place by two body scatterings.  The number
of quanta is conserved.  Following the logic through the isentropic expansion
stage, we see that the number of partons at formation is to a first
approximation the same as the number of produced pions.

A second issue concerns flavor production.  Initially most of the quanta are
gluons.  This is because they dominate the distribution functions.  The number
of quarks and anti-quarks in the sea is relatively small compared to the
number of gluons.  Therefore, the quark flavors come into chemical equilibrium
during the transport and hydrodynamic evolution times, and therefore can be
estimated by these methods if they turn out to be significantly in excess of
their intrinsic contribution to the hadron wavefunction.
                        
\section{Before the Parton Cascade}

In order to poperly formulate the initial value conditions for the parton
cascade, one must have a consistent quantum mechanical picture of the early
stages of the collision.  Such a picture is given by the McLerran-Venugopalan
model as extended to nucleus-nucleus collisions.\cite{mv}-cite{kmw}  The basic
ingredient in this
picture are non-abelian Lienard-Wiechart potentials.  To understand how this
works, consider an electric dipole at rest.  The electric field is the familiar
electric field shown in Fig. 2.

\begin{figure}
\begin{center}
\epsfxsize=10cm
\leavevmode
\hbox{ \epsffile{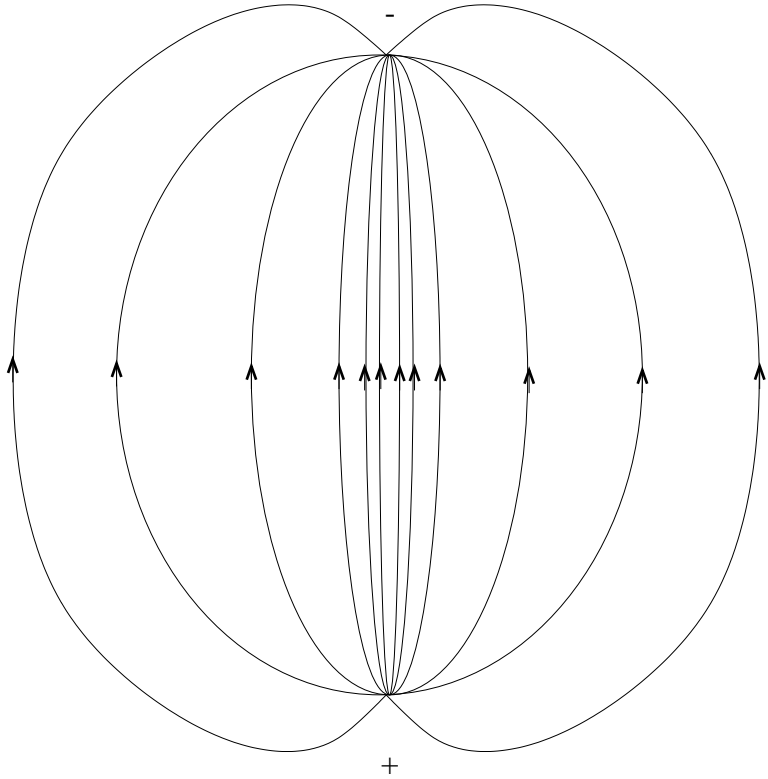}}
\end{center}
\caption{The Electric Field from a Dipole}
\label{fig2}
\end{figure}

If we now boost this field to the infinite
momentum frame, the electric and magnetic fields all exist in a plane
perpendicular to that of the direction of motion.  Further, the magnetic field lines
are perpendicular to the electric field lines.  Viewed head on, the electric field lines 
are those of Fig. 2, with magnetic fields everywhere orthogonal to electric.

Now if we study the field produced at central rapidity by a fast moving
nucleus, all the gluons at higher rapidity act as color sources for these
fields.  This means that the system is composed of very many dipole fields in
an infinitesmally thin plane perpendicular to the direction of motion.  Since
color is confined, on scales larger than that of a fermi, the fields vanish.
On smaller scales, they are stochastic.  The McLerran-Venugopalan model
assumes that these fields are generated by a Gaussian distribution of sources.
The weight function for these distributions of sources may be directly related
to the gluon distribution function.  

The fields maintain their Lienard-Wiechart form prior to the collision.  Upon
collision, the fields begin evolving.  The Yang-Mills equations can be solved
numerically from these initial conditions.\cite{kv}-cite{bmp}   Initially, the
fields are strong 
and the equations of motion are intrinsically non-linear.  As the fields
evolve, they dilute themselves and at some time the field equations linearize.
The solution to the linear equation corresponds to produced gluons.  One can
compute their phase space density, and this forms the initial conditions for a
subsequent cascade description.

There is only one scale in this classical problem:  the total charge in gluon
at rapidities other than the central region.  Up to powers of $\alpha_s$, this
is the same as the rapidity of gluon per unit area
\be
        \Lambda^2 = {1 \over {\pi R^2}} {{dN} \over {dy}}
\ee
If $\Lambda >> \Lambda_{QCD}$, then the coupling at this scale is weak, and
the classical description is consistent.  (Factors of $\alpha_s$ can be
ignored
in the power counting arguments below as they involve only logarithms of
density scales).

This single scale has many consequences.  It is precisely the scale-introduced
by hand in the parton cascade which is used to cutoff the parton cross
sections.  Note that it depends upon the initial density of partons per unit
area, and therefore upon rapidity and the baryon number of the target.
Verifying that there is in fact such a $p_T$ scale, and its dependence on
various nuclei and rapidity will be one of the things that RHIC should be able
to do.

Another consequence is because the density of produced gluon had the same
parametric dependence on density as does the initial gluon density per unit
rapidity, up to slowly varying factors of $\alpha_s$ and constant factors,
these densities are the same.  This provides an a posteriori justification for
the initial conditions used in the parton cascade model.  The space-time
structure of the initial conditions is automatically built in to the classical
computation. Several groups are now attempting solution of the classical field
nucleus-nucleus collision problems, and this can provide an initialization
for a parton cascade computation.

\section{Acknowledgments} 

I thank my colleagues Alejandro Ayala-Mercado, Miklos Gyulassy,
Yuri Kovchegov, Alex Kovner, Jamal Jalilian-Marian, Andrei Leonidov,
Raju Venugopalan and Heribert Weigert with whom the ideas presented
in this talk were developed.  This work was supported under Department
of
Energy grants in high energy and nuclear physics DOE-FG02-93ER-40764
and DOE-FG02-87-ER-40328.

\end{document}